\documentclass[aps,prl,twocolumn]{revtex4}
\usepackage{graphics}
\usepackage{graphicx}
\usepackage{epsf}
\usepackage{epsfig}
\usepackage{amsmath}
\usepackage{dcolumn}

\newcommand{\tx}[1]{\mathrm{#1}}

\begin{document}
\title{From Electrons to Finite Elements: A Concurrent Multiscale Approach for Metals}  
\author{Gang Lu$^{(1)}$, E. B. Tadmor$^{(2)}$ and Efthimios Kaxiras$^{(3)}$}
\affiliation
{
$^{(1)}$Department of Physics and Astronomy, California State University Northridge, 
Northridge, California 91330 \\ 
$^{(2)}$Department of Mechanical Engineering, Technion -- Israel Institute of Technology, 32000 Haifa, Israel \\ 
$^{(3)}$Department of Physics and Division of Engineering and Applied Sciences,\\
 Harvard University, Cambridge, Massachusetts 02138}
\date{\today}
\begin{abstract}
We present a multiscale modeling approach that 
concurrently couples quantum mechanical, classical atomistic and continuum 
mechanics simulations in a unified fashion for metals. 
This approach is particular useful for systems where 
chemical interactions in
a small region can affect the macroscopic properties of a material. 
We discuss how the coupling across 
different scales can be accomplished efficiently, and we apply 
the method to multiscale simulations of an edge dislocation 
in aluminum in the absence and presence of H impurities. 
\end{abstract}
\maketitle

Some of the most fascinating problems in all fields of science
involve multiple spatial and/or temporal scales: processes that
occur at a certain scale govern the behavior of the system across
several (usually larger) scales. In the context of materials science,
the ultimate microscopic constituents of materials are ions and 
valence electrons; 
interactions among them at the atomic level determine 
the behavior of the material at the macroscopic scale, the latter being the 
scale of interest for technological applications. 
Conceptually, two categories of 
multiscale simulations can be envisioned, 
sequential, consisting of passing information across scales, 
and concurrent, consisting of seamless coupling of scales \cite{review}. 
The majority of multiscale 
simulations that are currently in use are sequential ones,
which are effective in systems 
where the different scales are weakly coupled. 
For systems whose behavior at each scale 
depends strongly on what happens at the other scales, 
concurrent approaches are usually required.
In contrast to sequential approaches, concurrent simulations
are still relatively new and only a few models have been developed
to date \cite{review,qc,qm,maad,noam,lid}. 

A successful concurrent multiscale method is the Quasicontinuum (QC) method
originally proposed by Tadmor et al. \cite{qc}. 
The idea underlying this method is that atomistic processes of 
interest often occur in very small spatial domains while the vast 
majority of atoms in the material behave 
according to well-established continuum theories. 
To exploit this fact, the QC method retains 
atomic resolution only where necessary and grades out 
to a continuum finite element description 
elsewhere. The original formulation of QC was limited 
to classical potentials for describing interactions between atoms.
Since many materials properties depend {\it explicitly} 
on the behavior of electrons,  
such as bond breaking/forming at crack tips or defect cores, 
chemical reactions with impurities, 
and surface reactions and reconstructions, it is desirable to 
incorporate appropriate quantum mechanical descriptions 
into the QC formalism. In this Letter, 
we extend the original QC approach so that it can be 
directly coupled with quantum mechanical 
calculations based on density functional theory (DFT) 
for metallic systems. We refer to
the new approach as QCDFT.

The goal of the QC method is to model an atomistic 
system without explicitly treating every atom
in the problem \cite{qc,qc1}. This is achieved by replacing 
the full set of $N$ atoms with a small subset of $N_r$
``representative atoms'' or {\it repatoms} ($N_r\ll N$) 
that approximate the total energy through appropriate weighting. 
The energies of individual repatoms are computed in two different 
ways depending on the deformation in their 
immediate vicinity. Atoms experiencing large deformation 
gradients on an atomic-scale are 
computed in the same way as in a standard fully-atomistic method. 
In QC these atoms are called {\em nonlocal}
atoms to reflect the fact that their energy depends on the 
positions of their neighbors in addition to their 
own position. In contrast, the energies of atoms experiencing 
a smooth deformation field on the atomic scale are
computed based on the deformation gradient in their vicinity 
as befitting a continuum model. These atoms are 
called {\em local} atoms because their energy is based only 
on the deformation gradient at the point where it 
is computed. The total energy $E_{\rm tot}$ 
(which for a classical system can be written as 
$E_{\rm tot}=\sum_{i=1}^N E_i$, with $E_i$ the energy of atom $i$) 
is approximated as
\begin{equation}
E_{\rm {tot}}^{\rm {QC}}=\sum_{i=1}^{N^{{\rm nl}}}E_i(\{{\bf q}\})
+\sum_{j=1}^{N^{{\rm loc}}}n_jE_j^{{\rm loc}}(\{{\bf F}\}).
 \end{equation}
The total energy has been divided into two parts: an atomistic 
region of $N^{\rm nl}$ nonlocal atoms and a continuum region 
of $N^{{\rm loc}}$ local atoms 
($N^{{\rm nl}}+N^{{\rm loc}}=N^r$). The calculation in the 
atomistic region is identical to that in 
fully atomistic methods with the energy of the atom depending 
on the coordinates $\{{\bf q}\}$ of the
surrounding repatoms. However, in the coarse-grained continuum 
region each repatom can represent a large region of
$n_i$ atoms on the atomic scale. Rather than depending on the 
positions of neighboring atoms, the energy of a local 
repatom depends on the deformation gradients $\{{\bf F\}}$ 
characterizing the finite strain around its position. The basic 
assumption employed is the Cauchy-Born rule which relates 
the continuum deformation at a point to the motion of the atoms 
in the underlying lattice represented by this point. To obtain 
the necessary deformation gradients, a finite element mesh 
is defined with the representative atoms as its nodes. It is 
important to note that the calculations of 
$E^{{\rm loc}}_j(\{{\bf F}\})$ in the continuum regions make use 
of the same interatomic potential used in the nonlocal 
atomistic region. This makes the passage from the atomistic 
to continuum regions seamless since the same material description is 
used in both. This seamless description enables the model 
to adapt automatically to changing circumstances, 
for example the nucleation of new defects or the migration of existing defects. The adaptability of QC is one of 
its main strengths, which is missing in many other multiscale methods. 
A consequence of the partitioning into local and nonlocal 
regions and the existence of a well-defined
total energy for the entire system is the presence of 
non-physical {\it ghost forces} at the interface.
These can be eliminated by self-consistent application of 
dead load corrections \cite{qc1}.
	
\begin{figure}
\includegraphics[width=200pt]{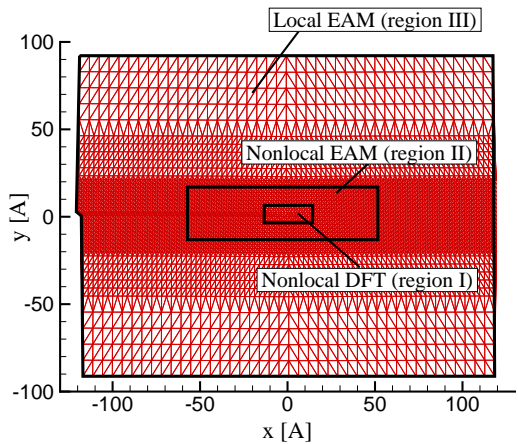}
\caption{QCDFT model for an edge dislocation showing the three 
domains in the model (see text). 
The x, y and z axes are along [$\bar{\rm 1}$10], [111] 
and [11$\bar{\rm 2}$], respectively. 
The model contains about 990 nonlocal atoms 
of which 84 are in the DFT region, 
3700 local atoms and 9300 finite elements.
All atoms are initially displaced according to the anisotropic 
elastic solution of the dislocation
with the boundaries held fixed to these values during the 
relaxation phase. Periodic boundary conditions
are applied along the dislocation line (z) direction.}
\label{fig:qcdft}
\end{figure}

The original QC formulation assumes that the total energy 
can be written as a sum over individual atom
energies. This condition is not satisfied by quantum mechanical models. 
To address this limitation,
in the present QCDFT approach the material of interest 
is partitioned into three distinct types of domains 
(see Fig.~\ref{fig:qcdft}): 
(1) a nonlocal quantum mechanical DFT region (region I); 
(2) a nonlocal classical region where Embedded-Atom 
Method (EAM) \cite{EAM} potentials are used (region II); 
and (3) a local region that employs the same EAM
potentials as region II (region III). 
The total energy of the QCDFT system is then
\begin{equation}
\label{energyqcdft}
E_{\rm {tot}}^{\rm {QCDFT}}=E[{\rm I+II}]
+\sum_{j=1}^{N^{{\rm loc}}}n_jE_j^{{\rm loc}}(\{{\bf F}\}),
\end{equation}
where $E[{\rm I+II}]$ is the total energy of regions I and II 
together (the assumption here is that region I is
embedded within region II). The coupling between regions II and 
III is achieved seamlessly via the 
QC formulation, while the coupling between regions I and II is 
accomplished by a scheme recently proposed by 
Choly {\it et al.} \cite{choly}. Based on this coupling strategy, 
$E[{\rm I+II}]$ can be written as 
\begin{equation}
\label{energy1}
E[{\rm I+II}]=E_{{\rm DFT}}[{\rm I}]+E_{\rm EAM}[{\rm II}]
+E^{\rm int}[{\rm I,II}],
\end{equation}
where $E_{\tx{DFT}}$[I] is the energy of region I in the absence 
of region II computed using the DFT model, 
$E_{\tx{EAM}}$[II] is the energy of region II in the absence of 
region I computed using the EAM model, 
and $E^{\tx{int}}$[I,II] represents a formal interaction energy 
added to give the correct total energy. 
The interaction energy between the two subsystems can be rewritten as:
\begin{eqnarray}
\label{array}
E^{\tx{int}}[{\rm I,II}] &\equiv& E[{\rm I+II}]-E[{\rm I}]-E[{\rm II}],
\\ \nonumber
& = &E_{\tx{EAM}}[{\rm I+II}]-E_{\tx{EAM}}[{\rm I}]-E_{\tx{EAM}}[{\rm II}].
\end{eqnarray}
The first equation serves as a general definition of the 
interaction energy whereas the second equation represents one 
particular implementation of $E^{\rm int}$, which is used in this work. 
Eq.~\ref{array} is not contradictory to 
Eq.~\ref{energy1} because EAM has its root in DFT and the EAM energy 
can be viewed as an approximation to the DFT energy. 
Different combinations of quantum mechanical and classical atomistic 
methods other than DFT/EAM may also be implemented \cite{choly}. 
The great advantage of the present implementation is its simplicity. 
It demands nothing beyond what is required for a 
DFT calculation and an EAM QC calculation. Furthermore, 
by substituting Eq.~\ref{array} into Eq.~\ref{energy1}, we arrive at
\begin{equation}
 \label{energy2}
E[{\rm I+II}]=E_{\tx{DFT}}[{\rm I}]-E_{\tx{EAM}}[{\rm I}]
+E_{\tx{EAM}}[{\rm I+II}].
\end{equation}       
The forces on the EAM atoms in region II are then
\begin{equation}
-{\bf F}^{\rm II}_i = 
\frac{E_{\rm {tot}}^{\rm {QCDFT}}}{\partial {\bf q}^{\rm II}_i}=
\frac{\partial E_{\tx{EAM}}[{\rm I+II}]}{\partial {\bf q}^{\rm II}_i}
+\frac{\sum_{j=1}^{N^{{\rm loc}}}n_jE_j^{{\rm loc}}(\{{\bf F}\})}{\partial {\bf q}^{\rm II}_i},
\end{equation}
where ${\bf q}^{\rm II}_i$ are the Cartesian coordinates of 
atom $i$ in region II. It is clear from this equation that
the forces on the atoms in region II are identical to those 
that would be obtained from a fully-classical QC calculation.
The same applies to the region III atoms, that is, 
as far as forces are concerned, regions II and III behave as 
though the entire model were classical. This is a very desirable 
property in terms of achieving a seamless coupling between 
region I and the rest of the model. At the same time,
the forces on the DFT atoms in region I will have contributions from 
both DFT atoms and the nearby EAM atoms in region II. 
The error in forces on the DFT atoms due to the coupling is thus 
given by the difference between calculated forces with DFT and EAM 
on these atoms. To minimize this error, we propose to use a 
class of interatomic potentials which are generated by 
matching the forces obtained from the EAM method 
to those from DFT calculations \cite{E-A,Ta}. Another important practical 
advantage of the present QCDFT method is that, if region 
I contains many different atomic species while region II 
contains only one atom type, there is no need to develop 
reliable EAM potentials that can describe each species 
and their interactions. This is because if the various species 
of atoms are well within region I, then the energy contributions 
of these atoms are canceled out in the total energy 
calculation (the last two terms in Eq.~\ref{energy2}). 
This advantage renders the method particularly useful in 
dealing with impurities, which is an exceedingly difficult 
task for empirical potential simulations. 

The equilibrium structure of the system is obtained by minimizing 
the total energy in Eq.~\ref{energyqcdft}
with respect to all degrees of freedom. Because the time required 
to evaluate $E_{\tx{DFT}}$[I] is considerably 
more than that required for computation of the other EAM terms 
in $E_{\rm {tot}}^{\rm {QCDFT}}$,
an alternate relaxation scheme turns out to be rather efficient. 
The total system can be relaxed by using the 
conjugate gradient approach on the DFT atoms alone, 
while fully relaxing the EAM atoms in region II and the 
displacement field in region III at each step.  
Similar to Choly {\it et al.} \cite{choly}, 
an auxiliary energy function can be defined as
\begin{equation}
\label{relax}
E'[\{{\bf q}^{\rm I}\}] \equiv \min_{\{{\bf q}^{\rm II}\},\{{\bf q}^{\rm III}\}}
E_{\rm tot}^{\rm QCDFT}[\{{\bf q}\}],
\end{equation}
which allows for the following relaxation scheme: 
(i) Minimize $E_{\rm tot}^{\rm QCDFT}$ with respect to the atoms in 
regions II ($\{{\bf q}^{\rm II}\}$) and the atoms in region III 
($\{{\bf q}^{\rm III}\}$), while holding the atoms
in region I fixed;
(ii) Calculate $E_{\rm tot}^{\rm QCDFT}[\{{\bf q}\}]$, 
and the forces on the region I atoms;
(iii) Perform a step of conjugate gradient minimization of $E'$;
(iv) Repeat until the system is relaxed. 
In this manner, the number of DFT calculations performed is greatly reduced, 
albeit at the expense of more EAM and local QC calculations. 
A number of tests have shown that the total number of DFT 
energy calculations for the relaxation of an entire system is 
about the same as that required for DFT 
relaxation of region I alone. Further computational speed-up 
can be achieved for the DFT calculations by using
converged electronic charge density and wave functions from the previous step, 
so that the charge (potential) self-consistency 
can be reached faster for the next DFT calculation because 
the atomic relaxation is usually very small between two 
consecutive DFT moves.

\begin{figure}
\includegraphics[width=180pt]{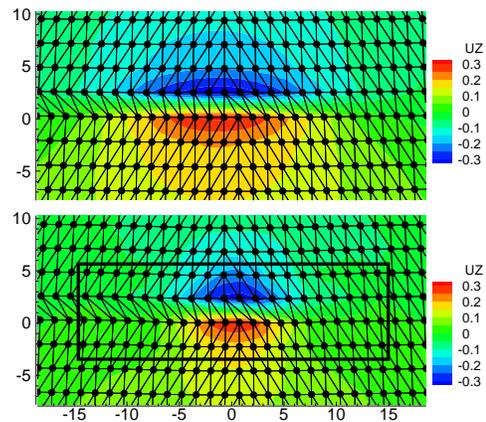}
\caption{Dislocation core structures obtained from the EAM-based QC 
(top) and the present QCDFT method (bottom).
The black circles are atoms. The contours correspond to out-of-plane 
(z) displacement in \AA. The contours clearly 
indicate the splitting of the dislocation. The atoms within the 
black box in the bottom panel are DFT atoms. 
The finite element mesh serves no other purpose in this nonlocal
atomistic region other than as a guide to the eye to help visualize deformation.}  
\label{fig:spread}
\end{figure}
In the remainder of the paper, we apply the present QCDFT 
approach to study the core structure of an edge 
dislocation in Al in the absence and presence of H impurities. 
We chose this system as an example because
results from both experiments and simulations are available 
for comparison. The QCDFT model for an
edge dislocation with a Burgers vector 
$\frac{a}{2}$[110] ($a=3.97$ \AA) is presented in Fig.~\ref{fig:qcdft}.
Convergence tests on the size of region I indicate that a 
DFT box of 30 \AA\ $\times$\ 9 \AA\ $\times$\ 4.86 \AA\ 
(84 DFT atoms) is sufficient to capture the dislocation 
splitting behavior accurately; hence the following calculations are 
all based on this DFT box. A force-matching potential for 
Al \cite{E-A} was used for EAM calculations. 
The DFT calculations were performed by using the plane-wave 
pseudopotential VASP code \cite{vasp} for a cluster with 8 \AA~ 
vacuum in both the x and y directions. The energy cutoff for pure 
Al and Al+H is 129 eV and 200 eV, respectively. We find 
that 10 $k$ points along the one-dimensional Brillouin zone are 
adequate for good convergence.  Fig.~\ref{fig:spread} presents the 
simulation results for both a standard EAM-based QC calculation and 
the QCDFT method, showing the dissociation of the edge 
dislocation into two equivalent 60$^\circ$ Shockley partials.
The splitting distance (obtained from an analysis of the
displacement jump across the slip plane) in the standard QC calculation 
is 15.4 \AA, whereas the splitting distance obtained with 
QCDFT method is 5.6 \AA, a value very close to the 
experimentally observed value of 5.5 \AA\ \cite{mills}. This result 
demonstrates that even for a simple metal like Al which 
should be the best candidate for use of an EAM potential, a 
quantum mechanical calculation is necessary to obtain correct results. 

\begin{figure}
\includegraphics[width=160pt]{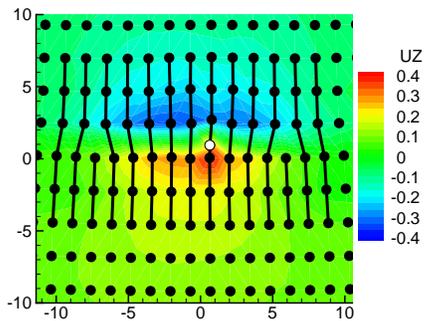}
\caption{QCDFT dislocation core structure in the presence of a column 
of H impurities. The circles are Al atoms (black) and H atom (white).
Contour significance is the same as in Fig.~\ref{fig:spread}. 
The black lines are a guide to the eye, indicating atomic planes.}
\label{fig:spread1H}
\end{figure}

The most important advantage of QCDFT approach, however, 
is that it allows the study of impurity effects on mechanical response, 
an impossible task for simpler empirical potentials. 
Fig.~\ref{fig:spread1H} shows the effect of adding one 
column of H impurities at the dislocation core. 
The presence of the H atoms results in a spreading of the core 
(the splitting distance is now increased to about 13 \AA). This finding 
is consistent with the fact, confirmed by earlier DFT calculations \cite{lu_h}, that H can lower the stacking fault energy. 
The fact that the dislocation becomes wider may explain the 
H-enhanced dislocation mobility that is 
believed to lead to H embrittlement phenomena via the 
enhanced local plasticity theory.  
A similar core structure is also found for two columns of H atoms 
placed at the dislocation core.
\begin{figure}
\includegraphics[width=160pt]{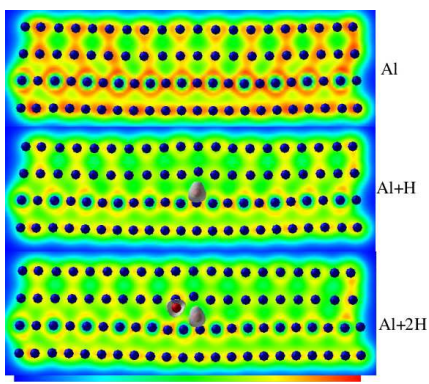}
\caption{Charge density distribution in region I in the absence (top) 
and in the presence of one 
(middle) and two H impurity atoms (bottom). 
The blue spheres are Al atoms and the red spheres are H atoms. 
The gray iso-surfaces illustrate
the charge density distribution at 0.28 electrons/\AA$^3$. 
Electron density values range from 0 to 0.30 electrons/\AA$^3$.}
\label{fig:charge}
\end{figure}
In order to understand the underlying origin of the H-enhanced dislocation 
mobility, we calculate the electron density distribution at the dislocation 
core in the absence and presence of H impurities, as shown in 
Fig.~\ref{fig:charge} . In the absence of the H impurity, the electron bonding 
is stronger and with a distinct covalent character. The bonding is more 
directional above the slip plane, and it becomes more spherical below the 
slip plane where there are two extra atomic planes, corresponding to the two 
partial dislocations. In the presence of H atoms, charge accumulation 
develops at these H atoms as the H impurities attract the valence electrons 
from the Al atoms and become negatively charged. The covalent bonding across 
the slip plane between Al atoms is disrupted by the H atoms, and at the same 
time, ionic bonding between the oppositely charged H and Al ions is developed. 
The fact that the directional covalent bonds are replaced by more homogeneous 
ionic bonds near the core leads to the wider dislocation core seen in 
Fig.~\ref{fig:spread}.

In summary, we have introduced a multiscale modeling approach which 
concurrently couples quantum mechanical, classical 
atomistic and continuum mechanics simulations, 
in a unified fashion for metals. 
Our QCDFT method provides a 
useful framework for multiscale modeling of metallic materials because 
it does not require the existence of localized
covalent bonds for computing the coupling energy as all  
other multiscale methods do \cite{maad,qm,noam,lid}. Furthermore,
this approach is completely general and versatile: 
it can be applied to diverse materials problems, such as
dislocations, cracks, surfaces, and grain boundaries. 
Finally, the automatic adaption feature of the QCDFT method 
allows the DFT and/or EAM region to move and change in response to 
the current deformation state, when for example,
defects are being nucleated in an otherwise perfect region. 
To demonstrate the unique strength of this method in dealing
with impurities, we have applied it to study H-dislocation interactions 
in Al.

\smallskip
This research was partly supported by an award from Research Corporation (GL).
ET and GL thank the Institute for Mathematics and its Applications (IMA) for 
hosting them in the fall of 2004 during which time part
of this work was done.

\end{document}